\documentclass[aps,prb,reprint,showpacs]{revtex4-1}
\usepackage[T1]{fontenc}
\usepackage{graphicx}
\usepackage[utf8]{inputenc}
\begin{document}

\title{The photoresponse of a two-dimensional electron gas at the second harmonic of the cyclotron resonance}
\author{M.~Bia{\l}ek}
\author{J.~{\L}usakowski}
\email{jerzy.lusakowski@fuw.edu.pl}
\affiliation{Faculty of Physics, University of Warsaw, ul. Pasteura 5, 02-093 Warsaw, Poland}
\author{M.~Czapkiewicz}
\affiliation{Institute of Physics, Polish Academy of Sciences, al. Lotnik\'{o}w 32/46, 02-668 Warsaw, Poland}
\author{J.~Wr\'{o}bel}
\affiliation{Institute of Physics, Polish Academy of Sciences, al. Lotnik\'{o}w 32/46, 02-668 Warsaw, Poland}
\affiliation{Faculty of Mathematics and Natural Sciences, Rzesz\'{o}w University, al.\ Rejtana 16A, 35-959 Rzesz\'{o}w, Poland}
\author{V.~Umansky}
\affiliation{Weizmann Institute of Science, Rehovot 76100, Israel}
\date{\today}
\begin{abstract}
Terahertz spectroscopy experiments at magnetic fields and low temperatures were carried out on samples of different gate shapes processed on a high electron mobility GaAs/AlGaAs heterostructure. For a given radiation frequency, multiple magnetoplasmon resonances were observed with a dispersion relation described within a local approximation of the magnetoconductivity tensor. The second harmonic of the cyclotron resonance was observed and its appearance was interpreted as resulting from a high frequency, inhomogeneous electromagnetic field on the border of a two-dimensional electron gas with a metallic gate and/or an ohmic contact. 
\end{abstract}
\pacs{52.35.Hr 73.21.-b 76.40.+b}
\maketitle

In a two-dimensional electron gas (2DEG) subject to a perpendicular magnetic field ($B$), electrical dipole transitions between Landau levels (LL) are restricted by the selection rule: $N \rightarrow N\pm1$, where $N=0, 1, ...$ is the LL number. However, this strict selection rule was shown to be relaxed in some cases. Observations of harmonics of the cyclotron resonance (CR) were reported in a Si metal-oxide-semiconductor field-effect transistor at a sufficiently high gate polarization.\cite{Kotthaus74} They were explained as a result of a simultaneous action of a short-range scattering  by charged traps at the Si/SiO$_2$ interface and the electron-electron scattering.\cite{Ando} Recently, Dai {\em et al.} reported on a strong response of a 2DEG at the second harmonic of the cyclotron resonance (2CR).\cite{Dai10} The experiments were carried out at temperature ($T$) between 0.3~K and 1.4~K on a GaAs/AlGaAs sample with a 2DEG mobility of $3\times 10^7$~cm$^2$/Vs and at
radiation frequency ($f$) from 60~GHz to 120~GHz. The 2CR peak was shown to be distinct from microwave-induced resistance oscillations  (MIRO)\cite{Zudov} and disappeared at about 1.4~K. A similar observation was reported by Hatke {\em et al.} on a 2DEG in a few GaAs/AlGaAs wafers with the electron mobility of the order of $10^7$~cm$^2$/Vs in experiments carried out at $T=0.5$~K or 1.5~K. In a series of papers \cite{Hatke11PRB83, Hatke11PRB84-121301, Hatke11PRB84-241304} properties of a peak appearing in the photoresponse in a proximity to the 2CR were analyzed in a great detail. It was shown that the peak did not result from the MIRO and that its properties could not be explained within existing theoretical models. 

In the present paper we report on the observation of a photoresponse  of a 2DEG at magnetic fields corresponding to the excitation of the 2CR. There are essential differences between our experiments and the experiments described in Refs.~\onlinecite{Dai10, Hatke11PRB83, Hatke11PRB84-121301, Hatke11PRB84-241304}. First, we investigated a 2DEG which resides in a single GaAs/AlGaAs heterostructure in comparison to modulation doped AlGaAs/GaAs/AlGaAs single quantum wells. Second, our experiments were carried out at 4.2~K and not in a mK regime. For this reason, the electron mobility in our samples was of the order of $6\times10^5$~cm$^2$/Vs and not $10^7$~cm$^2$/Vs. This clearly indicates that the conditions of our measurements did not correspond to that of the well developed MIRO. Third, we explored both the low (120--660~GHz) and high (1.5--2.5~THz) frequency bands, which essentially enlarges the frequency range used previously, and we show the 2CR response is observed at any of these frequencies.
Finally, we investigated both gated and ungated structures, in each case supplied with ohmic contacts, and we show that the presence of a metallic gate is crucial in generating a photoresponse at the 2CR, although there might happen cases when it is not the case.

We propose that the observed response at the 2CR is caused by a nonlinear cyclotron motion of an electron circulating in a strongly inhomogeneous electric field created by an incident radiation at a border of a metallic part of a sample (a gate or an ohmic contact) with a 2DEG. We discuss our results within predictions of two theories,\cite{Mikhailov11PRB83, Volkov14} both exploring a strongly inhomogenous high-frequency electric field at a border of a metallic contact and a 2DEG.

An important part of our reasoning is devoted to proving  that the observed photoresponse results from the 2CR and is not caused by avoided crossings Bernstein modes---magnetoplasmons interactions\cite{Bernstein58, Batke85, Krahne00, Kulik00}. This argumentation is important in the case of our results because excitations of magnetoplsasmons were found to be the main mechanism of the photoresponse observed in samples investigated.\cite{Bialek14APL, Bialek14JAP}

The samples used in the present study were processed on a high electron mobility GaAs/AlGaAs heterostructure grown by a molecular beam epitaxy.\cite{Bialek14JAP} Hall effect measurements at 4.2~K allowed us to estimate an electron concentration and a mobility to be about 3.0$\times$10$^{11}$~cm$^{-2}$ and 6$\times$10$^5$~cm$^2$/Vs, respectively. Samples were fabricated with an electron beam lithography. The processing started with a wet etching of a rectangular mesa (a typical dimensions was 65~$\mu$m $\times$ 1.3~mm). Next, ohmic contacts were fabricated by deposition of 160~nm of Au/Ge/Ni alloy and heating up to 430$^{o}$C for a few minutes. The Au/Pd gates were deposited directly on the sample surface. We investigated both gated and ungated samples. Gated samples differed by the gate shape; they were in the form of a rectangle, a meander or a comb. In the last two cases, the period of the gate structure was of the order of a few $\mu$m and the geometrical aspect ratio of about 0.5.
In each case, the gate covered all the distance between the ohmic contacs, with exception of 10~$\mu$m-wide slits adjacent to the contacts.  More details on the samples can be found in Ref.~\onlinecite{Bialek14APL}.  

Samples were placed in a liquid helium cryostat and cooled to 4.2~K with an exchange gas. A magnetic field, produced by a superconducting coil, was perpendicular to the sample surface. A THz radiation was guided to the sample with an oversized waveguide of the diameter of 12~mm. Two filters were used. A black polyethylene filter, placed next to the sample, filtered out a visible and a near-infrared light of the frequency higher than about 20~THz 
(including a room-temperature thermal radiation from warmer parts of the cryostat). A teflon filter stopped the radiation in the range from 15~THz to 40~THz. 
A THz beam was generated by an electronic source in the frequency range of 630--660~GHz, a backward-wave oscillator (BWO) generating at 100--170~GHz, 200--315~GHz, 315--485~GHz and a molecular THz laser pumped with a CO$_2$ laser (a few lines in the range 1.5--3.1~THz).

A source-drain photovoltage (PV) signal was registered as a function of a magnetic field at a constant radiation frequency. The photovoltage is generated by an incoming THz radiation only (without any bias of the sample). It appears due to an asymmetrical connection of the sample with one of ohmic contacts grounded and the other connected to a high-resistance input of a lock-in. A physical mechanism responsible for the observed photovoltage is a rectification of plasma excitations.\cite{Dyakonov96, Dyakonov05, Lifshits09} 
We used a phase-sensitive lock-in measurement system with an electronic on/off modulation of the radiation intensity. The modulation frequency was in the range of 13--500~Hz.
Typically, the signal amplitude was between a few $\mu$V up to 1~mV, depending on the radiation power coming to the sample and the magnetic field.

\begin{figure}
\includegraphics[width=8.5cm]{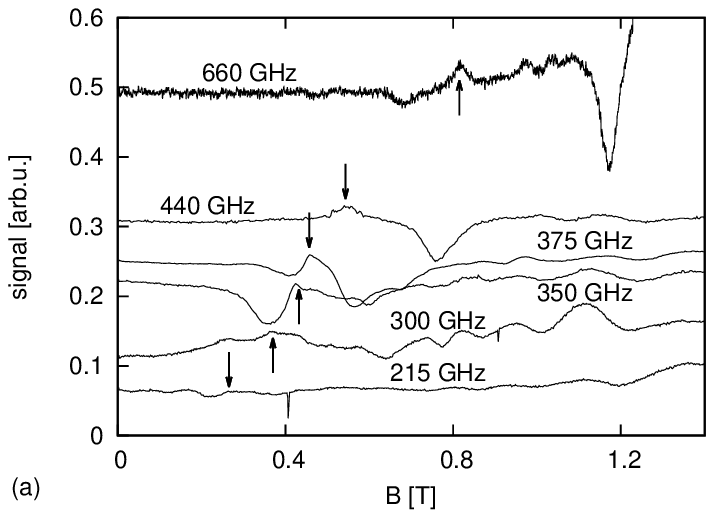}
\includegraphics[width=8.5cm]{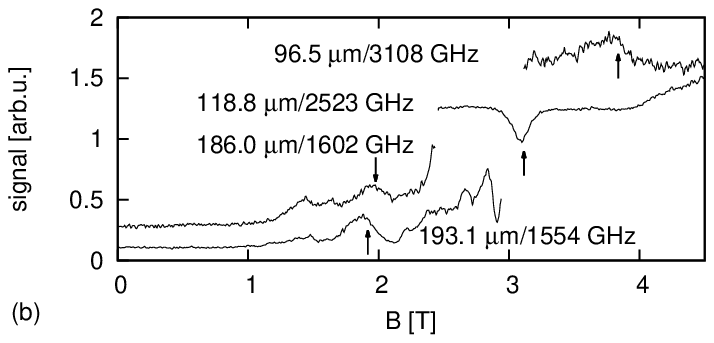}
\caption{Typical PV spectra collected for a meander-gated sample. Arrows show magnetic fields corresponding to the second harmonic of the cyclotron resonance. Results in (a) were obtained with the electronic source and the BWO, while in (b) with the THz laser.}
\label{spectra}
\end{figure}
\begin{figure}
 \includegraphics[width=7.5cm]{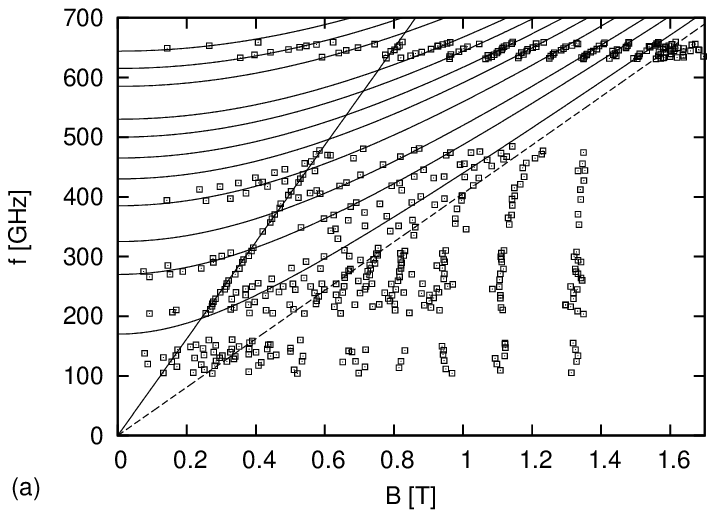}
 \includegraphics[width=7.5cm]{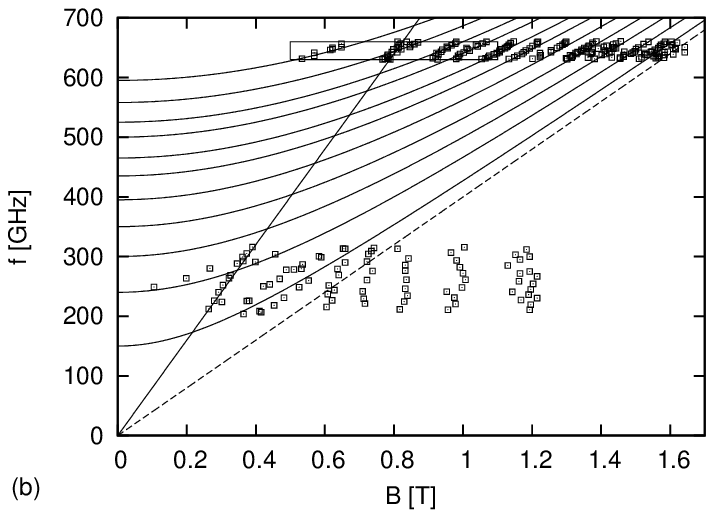}
 \includegraphics[width=7.5cm]{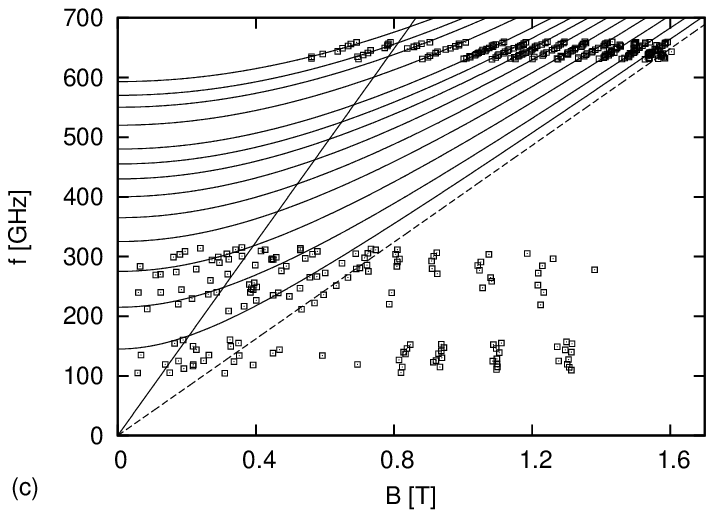}
 \caption{Positions of maxima in photovoltage spectra in the case of: (a) the meander-gated sample, (b) the uniformly-gated sample and (c) the ungated sample.}
 \label{points}
\end{figure}
Results obtained for a meander-gated sample are plotted in Fig.~\ref{spectra} which shows a PV as a function of the magnetic field in the frequency range of 200--700~GHz. At fields higher than the cyclotron resonance field $B_{CR}=2\pi f m^* / e$ (where $m^*=0.070m_e$ is the GaAs/AlGaAs heterostructure cyclotron mass, as determined from current experiment)\cite{Bialek14JAP} the dominant feature are Shubnikov-de Haas (SdH) oscillations.
Arrows mark positions of the second harmonic of the cyclotron resonance ($B_{CR}/2$). Other features visible in $B<B_{CR}$ fields are a superposition of magnetoplasmons excitations and SdH oscillations.\cite{Bialek14APL}

Fig.~\ref{points} shows positions of maxima in the spectra plotted as a  function of the radiation frequency $f$ used. The maxima change their positions following the relation for a magnetoplasmon frequency:
\begin{equation}
  \omega=\sqrt{\omega_c^2+\omega_{p,n}^2},
  \label{mp-eq}
\end{equation}
where $\omega=2\pi f$ is the magnetoplasmon (simultaneously, the radiation) frequency, $\omega_c=eB/m^*$ is the cyclotron frequency and $\omega_{p,n}$ is the $n$-th plasmon mode frequency at $B$=0. Using Eq.~\ref{mp-eq} we can fit the values of $\omega_{p,n}$, as shown in Fig.~\ref{points}. The dispersion relation of a plasmon at $B$=0 is given by:\cite{Stern67}
\begin{equation}
 \omega_{p,n}=\sqrt{\frac{Ne^2}{2m^*\epsilon_0}\frac{k_n}{\epsilon(k_n)}},
 \label{omega_p}
\end{equation}
where $N$ is the electron concentration,  and $k_n$ is a wave vector of $n$-th plasmon mode. The effective dielectric function $\epsilon(k_n)$ depends on whether the 2DEG is screened by the gate (gated):\cite{Eguiluz75}
\begin{equation}
  \epsilon_g(k_n)=\frac{1}{2}(\epsilon_1+\epsilon_2\coth(k_nd)),
  \label{epsg}
\end{equation}
or not screened (ungated):\cite{Popov05}
\begin{equation} 
  \epsilon_{ug}(k_n)=\frac{1}{2}(\epsilon_{1}+\epsilon_{2}\frac{1+\epsilon_{2}\tanh(k_nd)}{\epsilon_{2}+\tanh(k_nd)}).
  \label{epsug}
\end{equation}
From fitted values of $\omega_{p,n}$ we deduced a dispersion relation of plasmons. A detailed description of the procedure leading from measured spectra to the dispersion relations was presented in Ref.~\onlinecite{Bialek14APL}. What is important to a further analysis in the current paper---in the case of the meander- and the uniformly-gated samples---is that we found that wave vectors $k_n=(4n-1)\frac{\pi}{W}$ fit precisely to the theoretical dispersion with W=65~$\mu$m being the width of samples. In the meander-gated sample, plasmons excited due to the periodicity of the gate (8~$\mu$m) are also excited, but their frequencies are much higher than these of plasmons excited due to the width.\cite{Bialek14JAP} Calculations based on Eqs.~\ref{mp-eq} and~\ref{omega_p} show that only two grid-related magnetoplasmon modes could be observed in the spectral range used in the current experiment.
However, since their amplitude and width is similar to $W$-related magnetoplasmons they were not resolved in the spectra. 

\begin{figure}
 \includegraphics[width=8.5cm]{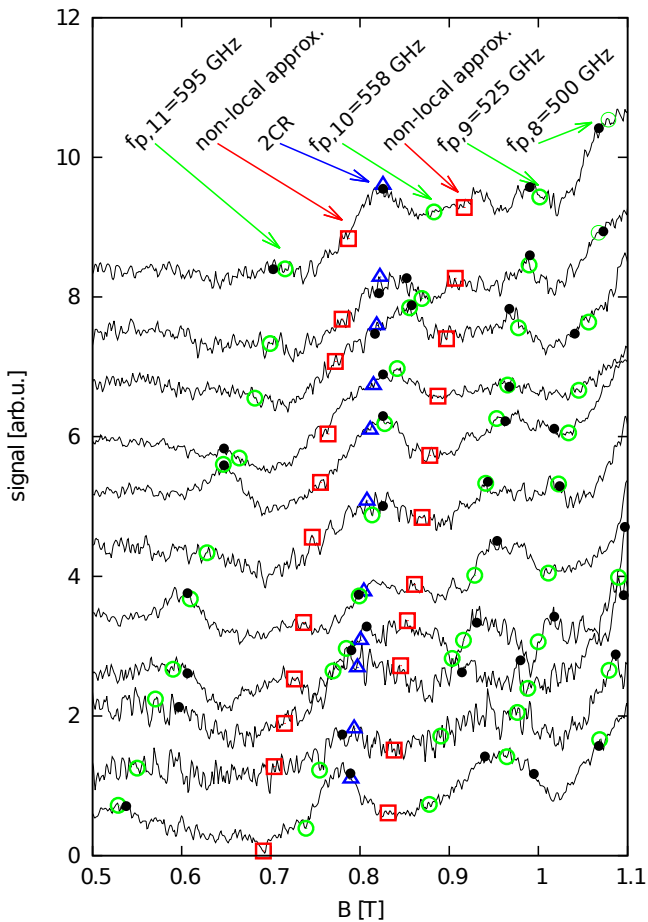}
 \caption{PV specta at the vicinity of the 2CR for the uniformly-gated sample. Results obtained in the 630--660 GHz range with a step of 2.88~GHz. Spectra are normalized and shifted for clarity. Maxima from these spectra are marked with black filled dots and are also shown in Fig.~\ref{points}(b) within a rectangle comprising a part of results obtained at above 600~GHz. Empty green dots, red rectangles and blue triangles show magnetic fields corresponding to theoretically calculated positions of the magnetoplasmons in a local approximation, 10th magnetoplasmon mode---Bernstein mode avoided crossing and the 2CR, respectively.}
 \label{waterfallH4}
\end{figure}

Let us note that a maximum occurring in the vicinity of the 2CR can be observed due to an interaction of a Bernstein mode\cite{Bernstein58} with magnetoplasmons. This interaction leads to a formation of avoided crossings in the magnetoplasmon spectrum which can be observed both in a bulk\cite{Wysmolek06} and 2D plasma.\cite{Batke85, Krahne00, Kulik00} In such a case, the interaction is characterized by the following features. First, consistently with an avoided crossing character of the interaction, corresponding maxima occur at magnetic fields close to but not equal to that of CR harmonics and an amplitude of a measured signal at the magnetic field of the 2CR should show a minimum in the middle of the interaction region.\cite{Batke85, Krahne00, Kulik00} Second, an amplitude of a Bernstein mode is comparable with that of a plasmon only within the interaction region and decaying rapidly it off.\cite{Platzman68, Horing1972, Batke85}

Basing on the experimental data presented in Figs.~\ref{spectra}--\ref{waterfallH4} we argue that a spectral feature observed at 2CR do not follow the above characteristics. A $B$-dependence of the 2CR is indicated in Fig.~\ref{points} with a solid straight line. As one can notice, there are series of points which coincide with the 2CR with a high accuracy and show a $B$-dependence clearly different from that of magnetoplasmons.  

Thus, the question is, whether these points correspond to the 2CR or to the mixed Bernstein mode---magnetoplasmons excitations. To answer this question, we show in Fig.~\ref{waterfallH4} a detailed analysis of a series of spectra in which we could observe an avoided crossing between one of the magnetoplasmon modes and the Bernstein mode close to the 2CR, if such an interaction were present. 
 
Inspection of spectra in Fig.~\ref{waterfallH4} show that at the magnetic field corresponding to the 2CR (marked with triangles) one systematically observes a single maximum of the signal which would not be the case if an avoided crossing of the Bernstein mode---magnetoplasmon were present. In the latter case, one would observe two maxima in the magnetic fields slightly higher and lower than the 2CR magnetic field and a minimum at the 2CR magnetic field.  A description of a magnetoplasmon dispersion within the local approximation is very accurate. A non-local approach is based on a solution of an equation:
\begin{equation}
 1-4\frac{\omega_{p,n}^2}{X^2}\sum_{m=1}^2\frac{m^2J_m^2(X)}{\omega^2-(m\omega_c)^2}=0,
\end{equation}
which describes the positions of excitations in the avoided crossing region of the interaction. 
Here $J_m$ is $m$-th order first kind Bessel function and $X=k_nv_F/\omega_c$ is a non-local parameter, where $k_n$ is equal to the $n$-th mode magnetoplasmon's wave vector and $v_F\approx2.2\times10^5$~m/s is the Fermi velocity. The $X$ value in our case is equal to about 0.15 for the 10th magnetoplasmon mode considered for the interaction. We cut the summation at the second term as we are considering only the interaction of a magnetoplasmon with the Bernstein mode coinciding with the 2CR. Parameter $\omega_{p,n}$ is the $n$-th plasmon mode frequency at zero magnetic field determined from fitting of Eq.~\ref{mp-eq} to maxima presented in Fig.~\ref{points}. Results of the calculations (open squares) are in a clear contradiction with the positions of maxima observed in the spectra. 
On the contrary, positions of the maxima in the spectra agree to a high accuracy with the positions of the 2CR and magnetoplasmons calculated within the local approximation. This allows us to argue that Bernstein modes are not observed in the present experiment. 
Finally, the 2CR feature was observed also in the measurements with the THz laser excitation (Fig.~\ref{spectra}(b)) at magnetic fields which are much smaller than the field at which magnetoplasmons were observed, which means that they are far off the interaction region. 

Since the 2CR peak was observed on all three gated samples and only on one among three ungated samples, we argue that it is generated due to processes that take place at a border of a 2DEG with a metallic part of a sample under an influence of a high-frequency THz radiation. Such a situation was considered in Refs.~\onlinecite{Mikhailov11PRB83, Volkov14} on which we base the interpretation of the results presented in this paper.
Essentially, both theories explore the fact that the incoming THz radiation generates a very strong and spatially nonuniform high-frequency electric field at a metal---2DEG border.

In Ref.~\onlinecite{Volkov14}, this field becomes particularly strong and oscillating in space when the radiation frequency approaches the double cyclotron frequency and falls within the band gap created by the Bernstein mode---magetoplasmon avoided crossing. Parametric resonance conditions give rise to a plasma instability causing a 2DEG heating, and therefore a~photoresponse at the 2CR frequency. An ultra high electron mobility, proportional to $\omega^{-3}$, is required to observe the signal at 2CR. According to the theory, we could see some 2CR signal at frequencies around 600~GHz in our samples. However, since the 2CR is observed at above 200~GHz, the required minimal electron mobility at this frequency should be about 15$\times$10$^6$~cm$^2$/Vs, which is of an order of magnitude more than the electron mobility in our samples.
Also, the theory predicts a photoresponse only in very narrow spectral regions in the gaps created by Bernstein modes---magnetoplasmon avoided crossings. We, however, observe the 2CR feature in a very wide spectral range. Reference~\onlinecite{Mikhailov11PRB83} describes effects originating from a penderomotive force, acting on electrons, present in a high-frequency nonuniform electric field at the vicinity of metallic contacts. An enhanced reaction to the incident radiation was predicted at magnetic fields $B_p$ satisfying $\omega=peB_p/m^*$, where $p$ is an integer. We propose that the structure discussed in the present paper corresponds to a $p=2$ feature. However, predictions of the theory about the shape of the 2CR peak do not agree with our experimental observations. We observe a single maximum instead of a maximum-minimum pair, the width of observed maximum is higher than predicted and appears to grow with the radiation frequency (Fig.~1). Thus, none of the theories fully explains a feature observed at the 2CR in our experiment.

In Refs.~\onlinecite{Mikhailov11PRB83,Volkov14}, a 2DEG was assumed to border a metallic contact which should be considered as an ohmic contact in a real experimental situation. 
The results of our experiments show that a metallic gate serves as better source of this nonlinearity. This should be not surprising because of a much sharper edge of a gate in a comparison with an edge of an ohmic contact which is typically slightly diffused. However, experiments show that an ohmic contact may be sometimes good enough, which was probably the case of one of our ungated samples. 

In conclusion, we presented an experimental evidence of a response of a 2DEG at the second harmonic of the cyclotron resonance. We propose that this is a manifestation of a complicated electron motion at the border of a 2DEG and a metallic gate or an ohmic contact due to a highly nonuniform electric field generated there by an incident radiation. 

Discussions with  S.~A.~Mikhailov,  M.~Dyakonov and K.~Nogajewski are gratefully acknowledged. This work was partially supported by a~Polish National Science Center grant DEC-2011/03/B/ST7/03062.

\bibliography{refs2}

\end{document}